# Blind High Dynamic Range Quality estimation by disentangling perceptual and noise features in images

Navaneeth Kamballur Kottayil, Giuseppe Valenzise, Frederic Dufaux, Irene Cheng, Anup basu

*Abstract*— No reference image quality assessment(NR-IQA) on true High Dynamic Range (HDR) images is an unexplored and interesting research topic that has become relevant with the current boom in HDR technology. We first conduct a study on the performance of existing algorithms for NR-IQA, originally developed for Low Dynamic Range images(LDR), on HDR data. Using the results of our investigation, we then propose a new convolutional architecture for NR-IQA achieving state of the art performance in terms of correlation to human judgment on quality. The proposed model disentangles perceptual effects exhibited by the distorted image from the noise present in the image and is capable of extracting perceptually relevant features without the necessity of handcrafting the features. Our architecture predicts the amount and location of the visual errors present in the distorted images without the reference image and performs comparably to the state of the art Full Reference Image Quality assessment(FR-IQA) algorithms.

## I. Introduction

The next wave of consumer imaging technology is in the field of high dynamic range imaging. A high dynamic range(HDR) image refers to an image that can capture a much larger range of luminance compared to a low dynamic range image. This translates to more vibrant colors, more detailed images and a better quality of experience of the end user of the content. HDR imaging technology is slowly becoming mainstream in the market and is reaching larger sections of people because of investments in the technology by camera and television manufacturers.

A problem of great academic and industrial interest is that of assessing the visual impact of the commonly seen image distortions in HDR images ( also known as image quality assessment or IQA). These arise mainly because of compression of the images by various compression schemes. Since the target audience for the HDR multimedia content a human being, the easiest way for IQA is through a subjective tests. However, these are often tedious and time-consuming. Even with massive crowdsourcing projects via systems like mturk, HDR IQA is hard because of expenses involved in acquiring systems capable of displaying the HDR content. Hence there is a need for high performance automated systems that are capable of IQA.

IQA can be mainly categorized into Full-Reference (FR), where the quality of the content is evaluated assuming that an undistorted content is available, and No-Reference (NR) where the quality of the content is evaluated considering that the undistorted content is not available.

While there is a rich amount of literature in the field of NR method of IQA of low dynamic range images (LDR) and for tone mapped images, not much exploration has been done in the area of NR-IQA for true HDR images (Luminances up to 4000 $Cd/m^2$) with luminance range greater than the conventional LDR display range (up to 300 $Cd/m^2$). In this publication we do a preliminary analysis of the performance of the existing LDR IQA algorithms exploring the problem of whether these algorithms can be used for HDR cases with minimal modifications.

We then use the analysis to propose a better model for HDR-NRIQA that is capable of predicting the image quality and localizing the location and amount of spatial error in an image. We use a convolutional neural network based architecture that computes perceptual features to derive the quality of an image.

The contributions of this paper are as follows: 1. A thorough study of NR quality prediction on HDR content. 2. Proposing methods to adapt NR-IQA for LDR on to HDR data. 3. Proposing an improved method for blind visual quality prediction in HDR images capable of producing results comparable to full reference image quality assessment algorithms 4. Providing an convolutional neural net architecture for separation of perceptual effects and noise in a distorted image.

For clarity of presentation, we divide the publication into two sections: A) Study of existing methods for NRIQA and analysis of performance on HDR data (section II) B) Proposed Method (Conceptual introduction in Section III, Implementation and results in Section IV and V).

## II. Performance of LDR No reference metrics on HDR data

In this section, we present the performance comparison study for LDR-NRIQA on HDR data and analysis of results.

There is no published work on NRIQA for true HDR images. There is, however, a lot of work on NRIQA for LDR images. Recent research has started looking into HDR images that has been tonemapped to LDR range by Tone Mapping Operators (TMO). The methods have not been adapted specifically for NRIQA on true HDR images. The methods developed for LDR images compare the statistical properties of an image and compare it with those of a natural image. The algorithms rely on the fact that image distortions alter the statistical properties exhibited by undistorted images.

Hence, the problem solved in this domain of research is that of modeling the change in statistics. The metrics model statistics of images by creating a feature image by using some processing method, and then fitting an arbitrary distribution on it. Followed by this, the parameters of this distribution is used as the feature vector of the distorted image (for example, DCT of the image would be utilized as a feature image and a Gaussian distribution would be fit on the same, the features would be the mean and variance of the Gaussian). The features are then used as inputs to some learning system, which is used to generate some score.

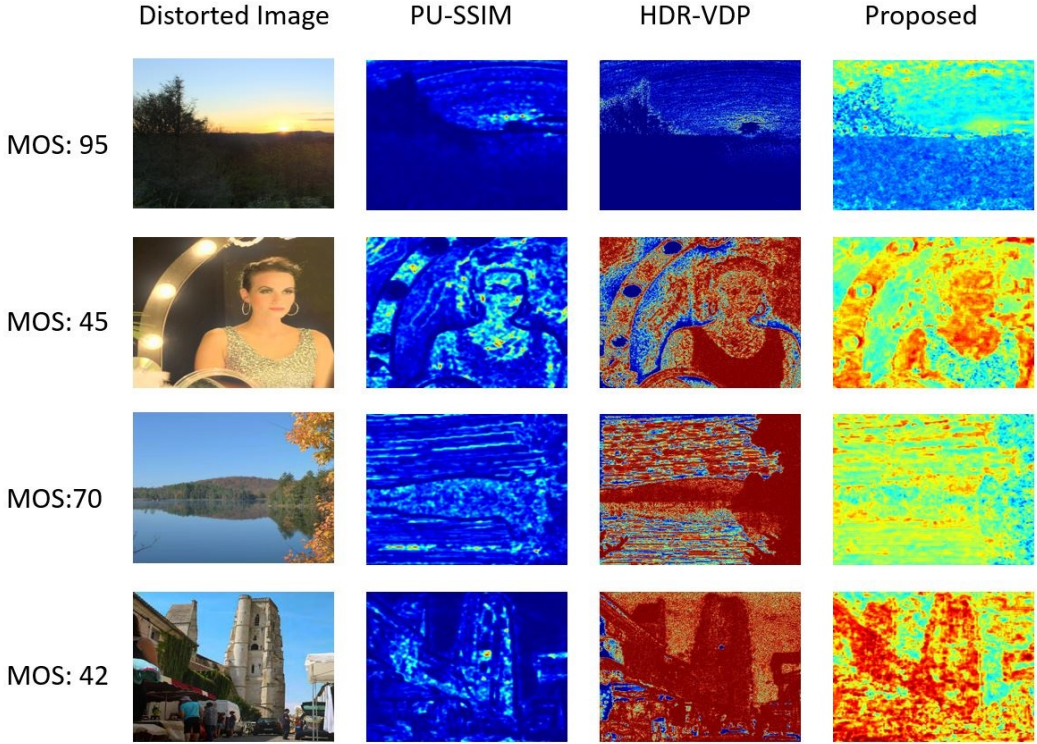

Fig. 1: Comparison of distortions in image estimated by various schemes. Red represents high value and blue represents a low value. Proposed method performs the estimation without the need for a reference image.

Most of the algorithms follow the process established by BIQI [13], which is a *two step process* where, from a set of features, an SVM predicts the type of distortion and another set of SVR's; one for each kind of distortion, predicts the score for each kind of distortion. The final quality scores are computed by

$$score = \sum_{i=1}^{m} p_i.q_i \quad (1)$$

where $p_i$ represents the probability of each distortion obtained from the SVM and $q_i$ represents the quality score given by each of the SVR's.

The feature image extracted varies with the method used. One of the first works in the field, BIQI [13], used Daubechies 9/7 wavelet as feature image. BRISQUE [12] computes a mean subtracted contrast normalized (MSCN) image as feature by using $MSCN(i,j) = \frac{I(i,j) - \mu_{I,N,i,j}}{\sigma_{I,N,i,j}+1}$, where $\mu_{I,i,j}$ and $\sigma_{I,i,j}$ represents the mean and variance computed over a local Gaussian window window of size $N$ around the point $i,j$. DIIVINE [14] used a divisive normalized steerable pyramid decomposition coefficients to create the feature image.

SSEQ [9] generates features using a different method. It uses entropy as features. Here, a scale space decomposition is carried out to generate three scales of images, and then entropy is calculated for image blocks in the spatial and DCT domain. This is then pooled by percentile pooling and the mean and variance of the spatial and frequency components are used as a feature vector for the two-step process.

An alternative approach that has the current state of the art results in LDR NRIQA is the Convolutional Neural Network (CNN) based approach used by Kang et al. , kCNN [4]. The publication uses a convolutional neural network with four layers on image patches of size 32X32. The first layer is a convolutional layer with 50 filters(kernels), then a pooling layer that reduces the dimensionality of the data and two fully connected layers, which is two sets of neurons with every neuron connected to every other neuron. The network is trained with the MOS scores. The method has an additional advantage that it can produce an 'error map' showing the visible errors on the distorted image.

From the literature, we understand that the degree of distortion is computed by some 'distance function' (generated by learning component in most cases) that measures how different the statistics of an image are from natural statistics. The natural statistics are not modeled explicitly but captured in the internal representation of the SVM or the CNN's used in the algorithms. Because of this, most of the proposed algorithms need an explicit training stage that adapts to the data before use. The training helps the learning component to understand what is a 'natural' feature and how the noise changes it.

## A. HDR and LDR statistics

The statistics of HDR images are quite different from that of LDR images. The fact was mathematically proved in the work done by Pouli et al. [17]. Hence the existing NRIQA cannot be directly used for HDR quality assessment. However, we hypothesize that the statistical differences between the two image types, however, will not impact the performance of NRIQA. We arrive at this hypothesis because of the learning system associated with NRIQA (SVM, CNN, etc.) for the prediction of the quality scores. Since the quality is predicted by the 'difference' of the image statistics of the image and a set of statistics internally stored by the SVM, the model can adapt to different statistics of the base images, as long as the noise statistics show a consistent pattern. Hence retraining NRIQA algorithms on HDR data will be the baseline for our comparisons.

## B. Perceptual factors affecting HDR data

As explained in the section above, the existing algorithms for NRIQA rely on image features to know what the natural statistic is and on a classifier to know how different the parameters of the input image is from that natural statistic. However, statistical differences from a 'natural' image might not be enough for a perceptual quality model for an HDR image. The reason for this stems from the fact that perceptual characteristics of HDR images are very different from that of LDR. This was extensively studied in Aydin et. al. [1]. The results indicate that given two images with the same type of distortion and with the same magnitude, a brighter display ( 1000 $cd/m^2$) causes a larger perceived distortion.

Hence, the study indicates a violation of Weber's law in HDR luminance ranges and suggests that a direct statistical comparison of the errors without considering these changes in sensitivity with luminance might not give the accurate model for a perceptual phenomenon.The authors of [1] proposed PU encoding as a solution to this problem to project the luminance on to a more perceptually uniform space where Weber's law is valid. This is a pre-processing step where HDR luminance values are converted to another range of intensity values.

We hypothesize that LDR-NRIQA algorithms can be easily adapted to HDR data by a PU encoding. The processing should theoretically simulate the various luminance adaptation effects seen in HDR scenarios and compensate for the effect. Once this aspect of the problem is solved, solving NRIQA reduces to that of computing the statistical effect of noise. Statistical modeling is already taken care of by the current LDR-NRIQA models.

## C. Tone mapping operators

Practically, PU encoding is implemented as a look-up table where one range of HDR luma values is converted into another range of values. The same operation is done by Tone Mapping Operators (TMO) as well; the difference between tone mapping and PU is just that tone mapping operators convert HDR luminance to LDR luminance range. It does not guarantee that the values after the operation are perceptually uniform.

We select a few highly cited methods to do tone mapping for our comparison. One of the first publications used the principles established in photography to change the local contrasts of the image. This was achieved in Reinhard et al. 2002[19], where a local luminance and contrast map was used to adjust the contrast of the image. The process was later simplified to produce a global scheme rather than local adjustment of contrast in Reinhard et al 2005[18]. Another popular scheme for tone-mapping is Drago et al [3]. The method was perceptually tuned and used a form of logarithmic compression of the intensity values. Mantiuk et al [10] proposed the use of an optimization over an error metric to produce an optimal tone mapping system. The system minimized the visible contrast distortions over the specified display type.

## D. Dataset

For a comprehensive dataset of true HDR image, we selected 5 separate data sets. The authors of the respective publications performed subjective evaluations using a SIM2 HDR47ES display with a maximum luminance of 4000 $Cd/m^2$. [15], consisting of JPEG compressed HDR images, [16] consisting of JPEG2000 compressed HDR images, [7] with JPEGXT compression and [21] containing images distorted by JPEG, JPEG2000 and JPEGXT compression schemes. The statistics of the data-sets are described in table I[[TODO: Wait for complete publication of Emin et al]] .

| Dataset Number | Number of Reference Images | Number of Distorted Images | Distortion type |
|---|---|---|---|
| #1 [15] | 27 | 140 | JPEG |
| #2 [16] | 29 | 210 | JPEG 2000 |
| #3 [7] | 24 | 240 | JPEG-XT |
| #4 [21] | 15 | 50 | JPEG JPEG2000 JPEGXT |
| #5 [?] | 15 | 50 | |

TABLE I: Database statistics

## E. Method

We test the performance of the existing NRIQA trained on LDR images on HDR dataset and after retraining the algorithms on HDR data by methods recommended in the respective publications. The features were extracted by the implementation provided by the authors. For training the SVM, methods suggested by the authors were used (SVR with RBF kernel). A grid search was conducted before the training to tune the hyper parameters of the SVM for optimal results.As per the original, we used 1000 iterations of training and testing, where 80% of images are used for training and 20% for testing, median scores of test cases reported.

The study of the effect of various preprocessing methods on performance was done using the same metrics by applying different pre-processing steps to HDR images. The results were obtained after retraining the algorithms on the respective processed HDR data. The pre-processing operators we choose are PU encoding and tone mapping using the schemes Reinhard 02 and 05 ([19], [18]), Drago [3] and Mantiuk [10].

*F. Results*

Comparison of performance was based on Spearman rank order correlation coefficient (SRCC), Kendall rank-order correlation coefficient (KRCC) and Pearson linear correlation coefficient (PLCC) and root mean square error (RMSE). A good NR-IQA is characterized by a higher value for SRCC, KRCC and PLCC and a lower value for RMSE. The performance of the LDR algorithms after simple retraining are given in Table II. The results of our study, ie performance in terms od SRCC, KLCC, and PLCC are shown in Tables II, III and IV.

Results from Table II, III and IV show a good performance after retraining the learning component of conventional NR-IQA. The method is adapting and capturing the statistics of the noise types even in HDR conditions. The relative differences in performance is similar to the ones observed in the case of LDR metrics. The state of the art performances are obtained for DIIVINE [14] and kCNN [5]. Practically, kCNN is more useful because it produces an error map in addition to the quality score. The error map shows an approximate error that is seen by the user on the noisy image.

From the performance after pre-processing the HDR values (Table II, III and IV), we observe a performance improvement in LDR-NRIQA algorithms if the data is pre-processed and the dynamic range of the data is reduced to LDR levels. As hypothesized in Section II.B, PU encoding seems to improve the performance for most of the cases. The maximum performance was obtained by the use of PU-encoded data with the kCNN. However, there are exceptions to this for example BRISQUE performs best with TMO by Mantiuk et al [10], BIQI best with TMO by Drago. The best performance is obtained in case of using PU in conjunction with kangCNN [4].

Our experiments above leads us to conclude the following A)LDR NRIQA methods by itself does not capture perceptually relevant features in HDR scenario. B) Adding a perceptual component that modulates the statistical noise measure (processing of images by PU encoding) improves performance in all cases C) PU does not seem to be optimal as there are cases where Tone mapping seem to have improved the results.

## III. PROPOSED METHOD

Designing a perceptual model that captures a large amount of interdependent image features interacting with each other is a challenging task. However, we have an understanding of the basic mechanism involved in the process. At its core, the problem can be represented as a set of two systems working with each-other. One detects the stimulus and another one that masks a certain part of this detected stimulus. Both these systems can be modeled without the need for rigidly handcrafting of their behaviors by the use of a convolutional neural network by constraining the network output functions.

To model the above idea, we implement a convolutional network architecture that acts on image blocks of HDR linear luminance values. We consider a block size of 32x32 pixels corresponding to the pixels in one visual degree under standard HDR test conditions. It consists of two systems acting on the data. The first part works on estimation of *noise* contribution $\delta(i,j)$ of the patch for a given image block centered at $i,j$. The second system computes the combined masking effect of the block; we refer to this as *error resistance* of block $T(i,j)$. These two terms are then mixed using a *mixing function* to produce the quality (we use Difference in Mean Opinion scores or DMOS) of the image block.

The proposed architecture is shown in Fig 2.

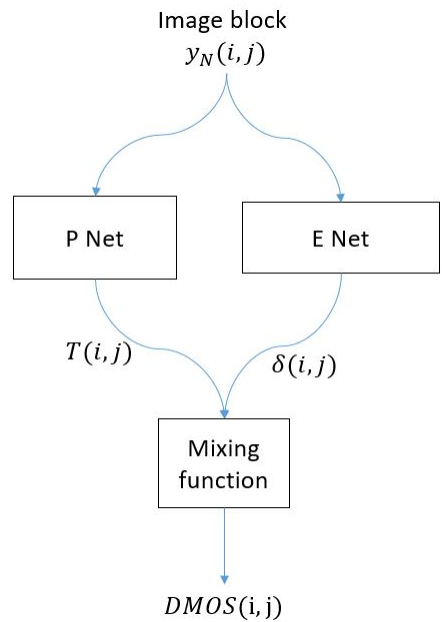

Fig. 2: Proposed strategy for IQA

The noise $\delta(i,j)$ represents the amount of unwanted statistics present in the image patch. For an image patch centered at $(i,j)$, we define the noise of the block as,

$$\delta(i,j) = mean(|Y_R(i,j) - Y_D(i,j)|) \qquad (2)$$

Where $Y_R$ and $Y_D$ represents the reference and distorted linear HDR luminance values of image block centered at the point $(i,j)$.

This is a problem that has already been tackled as part of the conventional LDR NRIQA techniques. And can be solved by a conventional CNN as it was demonstrated in [4]. Here, we use our own CNN to achieve this.

Next, for each patch centered at $(i,j)$, we compute a error resistance $T(i,j)$. This represents how resistant the block is to error. A high value for $T$ implies that a subject is less likely to see the noise $\delta(i,j)$ of the patch, hence the quality of the patch will be less affected by $\delta(i,j)$. A low value

| Processing | Linear | PU | TMO - Drago | TMO - Mantiuk | TMO - Reinhard 02 | TMO - Reinhard 05 |
|---|---|---|---|---|---|---|
| BRISQUE | 0.7942 | 0.8560 | 0.8272 | 0.8525 | 0.8306 | 0.6993 |
| SSEQ | 0.7891 | 0.8534 | 0.7999 | 0.8093 | 0.7962 | 0.6424 |
| BIQI | 0.8043 | 0.8012 | 0.8590 | 0.8368 | 0.8468 | 0.6709 |
| DIVIINE | 0.8815 | 0.8941 | 0.8507 | 0.8835 | 0.8586 | 0.7761 |
| kCNN | 0.8560 | 0.8980 | 0.8761 | 0.8947 | 0.8671 | 0.8052 |

TABLE II: SRCC performance of NRIQA with various preprocessing on HDR data

| Processing | Linear | PU | TMO - Drago | TMO - Mantiuk | TMO - Reinhard 02 | TMO - Reinhard 05 |
|---|---|---|---|---|---|---|
| BRISQUE | 0.6111 | 0.6787 | 0.6420 | 0.6670 | 0.6311 | 0.5179 |
| SSEQ | 0.6183 | 0.6864 | 0.6160 | 0.6203 | 0.6115 | 0.4645 |
| BIQI | 0.6208 | 0.6206 | 0.6778 | 0.6566 | 0.6605 | 0.4999 |
| DIVIINE | 0.7227 | 0.7277 | 0.6782 | 0.7100 | 0.6792 | 0.5874 |
| kCNN | 0.6678 | 0.7176 | 0.6881 | 0.7177 | 0.6702 | 0.6180 |

TABLE III: KLCC performance of NRIQA with various pre-processing on HDR data

| Processing | Linear | PU | TMO - Drago | TMO - Mantiuk | TMO - Reinhard 02 | TMO - Reinhard 05 |
|---|---|---|---|---|---|---|
| BRISQUE | 0.7825 | 0.8391 | 0.8190 | 0.8432 | 0.7995 | 0.7048 |
| SSEQ | 0.7772 | 0.8432 | 0.7970 | 0.7934 | 0.7961 | 0.6489 |
| BIQI | 0.7909 | 0.7863 | 0.8504 | 0.8263 | 0.8391 | 0.6774 |
| DIVIINE | 0.8833 | 0.8869 | 0.8473 | 0.8774 | 0.8512 | 0.7727 |
| kCNN | 0.8441 | 0.8735 | 0.8479 | 0.8626 | 0.8466 | 0.7807 |

TABLE IV: PLCC performance of NRIQA with various pre-processing on HDR data

implies that the image patch will be perceptually degraded by noise.

Error resistance $T(i,j)$, represents the combined the perceptual effects like luminance adaption and spatial masking exhibited by the block. It can be considered as similar to the pixel wise error detection thresholds in conventional IQA systems like [2] and [11]. These were determined by mathematical modeling of the response of the HVS to the region surrounding $(i,j)$. Traditionally called as contrast sensitivity model, they consider various visual features like contrast frequency etc into account and use a custom function to mix all these factors together. The functions in turn were derived from various psycovisual experiments. The experiments are often based on HVS response data to sinusoidal gratings. It cannot be assumed that these methods can generalize to a real world image with millions of frequencies and luminance and contrast levels.

We tackle this issue by the use of a convolutional network based architecture to derive the error resistance of the block. The method would compute the features required to do this task from the real world image data provided to it by numerical optimization. This provides the guarantee that the threshold work on real world images; hence instead of manually determining the functions that determine how the error resistance is computed, we assign a neural network to select from a potentially large number of 'functions'(that are represented as neural network weights) and provide the best one suited for the observed the data.

We combine the two values with a *mixing function*, represented as $f(\delta, T)$. This is an important part of the formulation since it determines the behavior of P-net. The final result is being optimized by the training process to match quality score, hence depending on this function the output of P-net would change. For example, if we choose a simple product as the mixing function (ie $DMOS = T*\delta$) and the network converges successfully, P-net would theoretically produce an overall scaling factor that changes according to how sensitive the region is to errors. Similarly, if a division is used and the other issues discussed are met, a factor similar to error detection thresholds would be produced by P-net.

While it can be argued that this also can be left to another convolutional neural network to decide, such a formulation would involve more weights and difficulties in optimization and could result in the optimization not converging to a good solution. Additionally, this function would be a 'black box' with no intuitive interpretations.

An intuitive way to simplify the training and guarantee convergence and interpretability would be to use a function that already exhibit sigmoid like behaviour and then optimize the network. This would also make the model more intuitive and easier to analyze in the context of our the existing knowledge on perceptual systems. For two cases product and division as mixing function described above, the network would have to learn the sigmoid like behavior exhibited by the human visual system when detecting changes in stimuli. Practically, this translates to more complex features being learned in the weights and a more difficult optimization. Note that choosing a function that is too complex can also lead to the same optimisation problems because of unstable points along the function or low values for gradients leading to slow/zero learning etc. We donot go into the mathematics of this as it is beyond the scope of this work.

We tackle this problem by using the nature of the statistics that we see in pyscovisual experiments; we expect a function of the two proposed variables representing DMOS of the image patch to be monotonically increasing with $\delta$ and decreasing with $T$. We choose a hyperbolic tan function since the function models the perceptual characteristics we want and is the activation function conventionally used in neural networks([8]); hence there are no issues with the convergence.

$$DMOS(i,j) = tanh(\frac{K * \delta(i,j)}{T(i,j)})  \quad (3)$$

As seen from the plot of equation 3 in figure 3, the function would represent a wide range of values and rate of increase, depending on the value of $\delta(i,j)$, $T(i,j)$ and $k$. This models the expected trends in DMOS for different values of error and error resistance.

For example, for the error resistance $T(i,j)$ that is large, the output of the function would remain small even if there is a large noise $\delta(i,j)$; case $T = 200$ in fig 3. This can be further modulated by varying the variable $k$. This scenario would be seen in cases of high masking because of luminance or complex patterns. Conversely, if the error resistance is low, the predicted DMOS would shoot up very quickly even for the smallest of noise. Practically, this would be seen in areas that are smooth with very low masking effects.

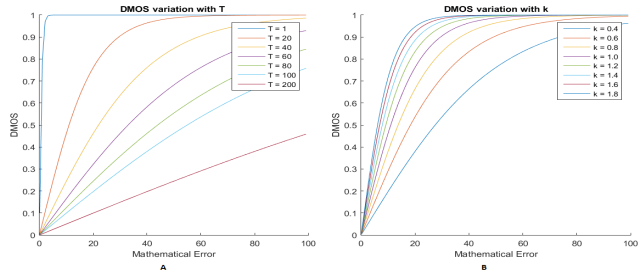

Fig. 3: Behavior of the mixing function. (A) Varying T with k fixed at 1. (B) Varying k with T fixed at 20.

The behavior of the function is very similar to that of a conventional sigmoid function based error detection model dependent on an error value and a error resistance mechanism in publications like [22] and [11].

A conventional approach of deriving the degree of distortion or error probability that would require a weighted pooling scheme like minkovski pooling or saliency weighted pooling. We average the patch scores across the whole image. In our case a simple averaging would produce a good result because the patch scores themselves represent the local DMOS scores.

## IV. DESIGN

The proposed network architecture for the Error estimation (E Net) has 5 layers. E Net is required to do a blind noise estimation. Hence we choose a typical CNN architecture consisting of 5 layers. The layers are convolutional with 64 filters of dimension 7X7, 128 filters of 5X5, 256 filter of 3X3 and 512 1X1 filters. Spatial pooling of 2X2 was used after each filtering stage. The final layer consisted one node corresponding to the output. Spatial dropout layers [20] were added to prevent over-fitting of the data.

P Net is required to estimate the error resistance values of the block. Here, we define a custom CNN layer that we call as the Augmented input layer with additional information. In addition to the original luminance values of the patch, we compute the variance and MSCN images. The variance image is computed by replacing every pixel $(i,j)$ with variance computed over a local Gaussian window window of size $N$ around the point $i,j$. For MSCN image, we use the equation proposed in [12], $MSCN(y_N(i,j)) = \frac{y_N(i,j) - \mu_{y_N(i,j)}}{\sigma_{y_N(i,j)} + 0.01}$. $\mu_{y_N(i,j)}$ is computed by replacing every pixel $(i,j)$ with the mean computed over a local Gaussian window window of size $N$ around the point $i,j$; We use a smaller value for the stabilizing constant to prevent the stabilizing constant from influencing the MSCN values. Since a neural network training requires that the input value be in a similar range, we scale the input, variance map and the MSCN map with a trainable weight whose values are determined as part of the overall optimization process. Hence the output of the augmented layer will be $[<W_1*y_N(i,j)>, <W_2*\sigma_{y_N(i,j)}>, <W_3*MSCN(y_N(i,j))>]$.

The following layers consist convolutional layers consisting of 64 filters of dimension 3X3 and 128 filters of 3x3. Followed by this 2 densely connected layers with 100 nodes each. The final layer has one node corresponding to the output.

The network structure is shown in figure 4 and 5.

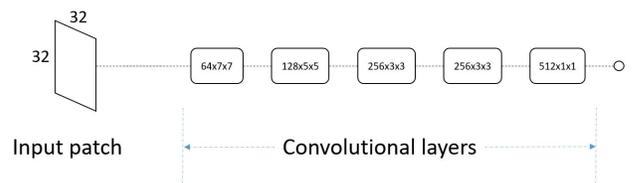

Fig. 4: Network structure for Error estimation.

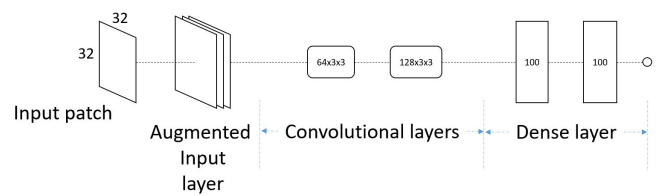

Fig. 5: Network structure for error resistance value.

The results of the two networks pass through another custom mixing layer whose behavior can be modeled by equation 3. Here the parameter $k$ is tuned as part of the training process. We use Adam optimizer with parameter values for learning rate=0.001, $\beta_1$=0.9, $\beta_2$=0.999, $\epsilon$=1e-08, decay=0.0 for training the CNN (as per recommendations in the original publication [6]).

### A. Training

The biggest drawback on a CNN based system is the large amount of training data required to compute the weights of the neural network.

E-Net estimates noise; the training data for this task can be easily obtained. For each patch of the image, the target value would be the mean error in that patch. This in turn is the difference between the distorted and reference images (Equation 2).

For training P-Net, the ideal training data would be a number that combines all the perceptual effects of the human visual system acting on the image patch. We donot have this data, however, we can obtain value for the final patch quality after the mixing function. We use this for training.

In training P-net, we make a strong assumption that the patch quality is equal to the global image quality score. Even though this assumption is incorrect most of the time, it was shown in [5] that, with a starting assumption that the global quality of the distorted image is the same as that of the local quality, the training process of the CNN isolates the local quality. The publication by Kang et al. [5] found that the CNN detected spatial patterns of noise in the feature image provided. Under the assumption we make, multiple quality scores might be associated with the same pattern of noise, however, when trained over millions of patches with a cost function that imposes sparcity constraint (the paper used L1 distance between the predicted quality and the actual quality), the correct local quality is the only value that minimizes the total error. In other words, the lowest cost of the CNN cost function was obtained when the CNN generated the true local errors of the image, regardless of the labels it started out with.

Thus define our *two stage training process*; In stage 1, E-net is trained with image patches as input and the corresponding mean error of the image patch as target, hence E-Net learns the patterns in the data corresponding to noise.

Then, in stage 2, all the training weights of E-net is frozen by dropping the learning rate of this section to zero. The whole network is then trained with image patch as input and global image quality score as target. The process is illustrated in 6.

| Feature | Processing | SRCC | KLCC | PLCC | RMSE |
|---|---|---|---|---|---|
| brisque | Lin | 0.7942 | 0.6111 | 0.7825 | 1.1021 |
| brisque | PU | 0.8560 | 0.6787 | 0.8391 | 0.9784 |
| brisque | TMO - Drago | 0.8272 | 0.6420 | 0.8190 | 1.9397 |
| brisque | TMO - Mantiuk | 0.8525 | 0.6670 | 0.8432 | 1.2279 |
| brisque | TMO - Reinhard 02 | 0.8306 | 0.6311 | 0.7995 | 1.1801 |
| brisque | TMO - Reinhard 05 | 0.6993 | 0.5179 | 0.7048 | 2.0697 |
| sseq | Lin | 0.7891 | 0.6183 | 0.7772 | 1.2401 |
| sseq | PU | 0.8534 | 0.6864 | 0.8432 | 1.8855 |
| sseq | TMO - Drago | 0.7999 | 0.6160 | 0.7970 | 1.2111 |
| sseq | TMO - Mantiuk | 0.8093 | 0.6203 | 0.7934 | 1.0453 |
| sseq | TMO - Reinhard 02 | 0.7962 | 0.6115 | 0.7961 | 1.1505 |
| sseq | TMO - Reinhard 05 | 0.6424 | 0.4645 | 0.6489 | 1.4053 |
| biqi | Lin | 0.8043 | 0.6208 | 0.7909 | 1.2681 |
| biqi | PU | 0.8012 | 0.6206 | 0.7863 | 1.1565 |
| biqi | TMO - Drago | 0.8590 | 0.6778 | 0.8504 | 0.8700 |
| biqi | TMO - Mantiuk | 0.8368 | 0.6566 | 0.8263 | 1.0884 |
| biqi | TMO - Reinhard 02 | 0.8468 | 0.6605 | 0.8391 | 1.1248 |
| biqi | TMO - Reinhard 05 | 0.6709 | 0.4999 | 0.6774 | 1.5594 |
| diviine | Lin | 0.8815 | 0.7227 | 0.8833 | 0.9455 |
| diviine | PU | 0.8941 | 0.7277 | 0.8869 | 1.0823 |
| diviine | TMO - Drago | 0.8507 | 0.6782 | 0.8473 | 1.4253 |
| diviine | TMO - Mantiuk | 0.8835 | 0.7100 | 0.8774 | 1.1238 |
| diviine | TMO - Reinhard 02 | 0.8586 | 0.6792 | 0.8512 | 1.4639 |
| diviine | TMO - Reinhard 05 | 0.7761 | 0.5874 | 0.7727 | 1.3315 |
| kCNN | Lin | 0.8560 | 0.6678 | 0.8441 | 1.3892 |
| kCNN | PU | 0.8980 | 0.7176 | 0.8735 | 1.1982 |
| kCNN | TMO - Drago | 0.8761 | 0.6881 | 0.8479 | 1.6288 |
| kCNN | TMO - Mantiuk | 0.8947 | 0.7177 | 0.8626 | 1.5331 |
| kCNN | TMO - Reinhard 02 | 0.8671 | 0.6702 | 0.8466 | 1.6554 |
| kCNN | TMO - Reinhard 05 | 0.8052 | 0.6180 | 0.7807 | 1.8663 |
| **Proposed** | **Lin** | **0.9164** | **0.748**1 | **0.9090** | **1.077** |

TABLE V: Overall Performance comparison

## V. RESULTS

The algorithm was implemented on a moderately powerful computer with a Intel core i7 processor, 16GB RAM and an Nvidia GTX660 graphics processor. The language used was python with theano and open CV as supporting libraries.

### A. Performance Evaluation

For measure of performance, we use the same metrics as before ie, Pearson correlation coefficient (PLCC), Spearman rank-order correlation coefficient (SROCC), Kendall correlation coefficient (KLCC) and Root mean squared error (RMSE). A larger value for SCRR, PLCC and KRCC and a lower value of RMSE is expected to be attributed to a better performing metric.

The testing process also remains the same as before ie 1000 iterations of training and testing, where 80% of images are used for training and 20% for testing, median scores of test cases reported. The performance of all the chosen algorithms and the proposed algorithm on the combined data set is shown in Table V. It is clear from the test cases that the performance of the proposed system is better than the other algorithms even after compensation with PU encoding.

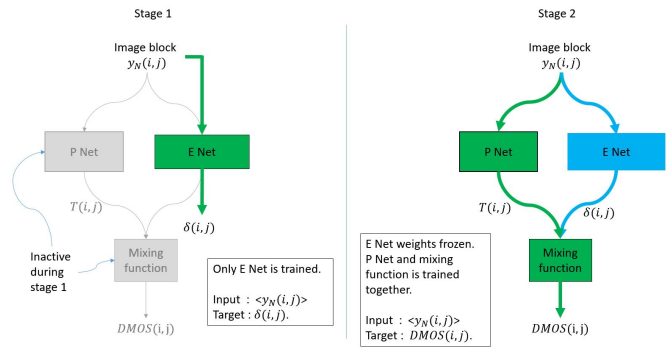

Fig. 6: Two stage training Process.

This process forces the P-net to extract a set of features from the image patch and derive a single error resistance value from it. When this value is combined with the output of E-net using the mixing function produces the required quality score of the image patch.

### B. Real world test

One of the common complaints against an NRIQA system is that the performances donot generalize with a different content. To test for this, we train the algorithms using datasets #1,#2 and #3 and test it on #4 and 5. This represents a real world scenario where, there is a completely different content with different image size as that of the training data.

In addition to this, this method of testing also allows us to perform a head to head real world test against the performance of full reference image quality assessment algorithms. From a machine learning point of view, this is acceptable, since we have sufficient number of examples of each type of distortion in datasets #1,#2 and #3 and a combination of all of the distortions in dataset #4 and 5. The test set contains DMOS scores uniformly distribute in the range [20,80].

Since the CNN are initialized with a random set of weights, the results of training can vary. We report the median score after 3 train test cycles. Our results for real world test is given in VI.

The results here are very interesting. The most notable fact is that the BRISQUE,BIQI,SSEQ and DIIVIINE seem to be unable to adapt to the different image sizes in the dataset if it is not trained. This can be explained by the fact that the features are computed over a joint histogram from the entire image.

The CNN based method performs well and shows good adaptability to a different test case. This can be attributed to the fact that an image patch is used to train the CNN and hence the overall image size becomes less of a factor. The features that make a difference here would be the patterns in quantification artifacts within each on the image patch.

Our method is generalizing very well to a difference in image content achieving performance very close to full reference algorithms, though there is still room for improvement here. A scatter plot of the scores produced by the proposed method to actual DMOS is shown in 7 low and medium brightness with slow variations would be attributed to lower masking value hence, would attribute to less masking, higher noise visibility and less error resistance.

| Feature | Processing | SRCC | KLCC | PLCC |
|---|---|---|---|---|
| brisque | Lin | 0.7194 | 0.537 | 0.7094 |
| brisque | PU | 0.4890 | 0.3643 | 0.5363 |
| sseq | Lin | 0.7203 | 0.5240 | 0.7043 |
| sseq | PU | 0.6331 | 0.4416 | 0.6109 |
| biqi | Lin | 0.3801 | 0.2577 | 0.3603 |
| biqi | PU | 0.2158 | 0.1462 | 0.2026 |
| diviine | Lin | 0.4362 | 0.3025 | 0.3938 |
| diviine | PU | 0.5208 | 0.3623 | 0.5007 |
| kCNN | Lin | 0.6732 | 0.4869 | 0.5133 |
| kCNN | PU | 0.7369 | 0.5390 | 0.7244 |
| HDR-VDP-2.2 | Full Reference | **0.9247** | 0.520 | **0.9407** |
| HDR-VQM | Full Reference | 0.9196 | 0.530 | 0.9333 |
| PU-MSSIM | Full Reference | 0.8969 | 0.570 | 0.9038 |
| **Proposed** | Lin | 0.8779 | **0.6935** | 0.8907 |

TABLE VI: Real world performance

To test this, we generate a simple modulated sinusoidal grating of dimension 800x800, with luminance value scaled to have maximum value of 4000, shown in fig 8. We then generate alternate versions of this with different luminance scale factors, obtaining a range of maximum luminance ranges.

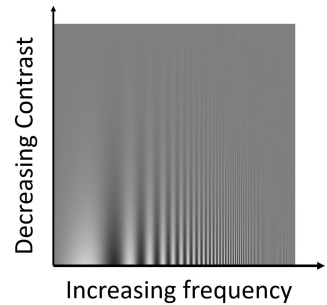

Fig. 8: Image for testing error resistance.

We give this image as input to the P-net (as patches of size 32x32) and examine the output images. The results are shown in fig 9.

The trends observed here are as expected, consider any one error resistance map of one luminance range (say Error resistance corresponding to image with luminance range [0,4000] ie row 2, image 4). The value of the error resistance value increases along x axis with frequency to a point and then decreases, indicating a trend in sensitivity to noise. This can be explained by observing the images themselves. We can perceive the high frequency variations in pixels of the original image till one point. Hence any noise embedded here would be masked because of spatial masking by the perceived frequencies. This is indicated by the higher values of the error resistance value. After this point, our ability to perceive changes in frequency falls and the region in original image appears smooth (rightmost side of row 1 image 4), any noise added to this point would become visible again; as indicated by the low error resistances of the corresponding

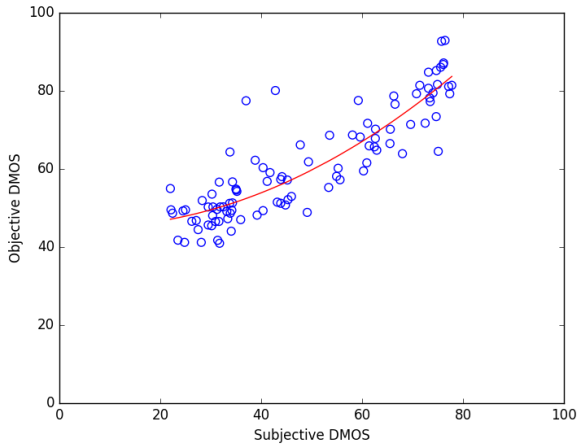

Fig. 7: Scatter plot between objective scores by proposed method and actual DMOS

## C. Error Resistance

The perceptual component of the architecture we proposed is the error resistance produced by P Net. We expect this value to change depending on the image data given; large value of luminance and high frequency would generally have a larger masking effect leading to low noise visibility and hence we expect the error resistance to be high. Conversely,

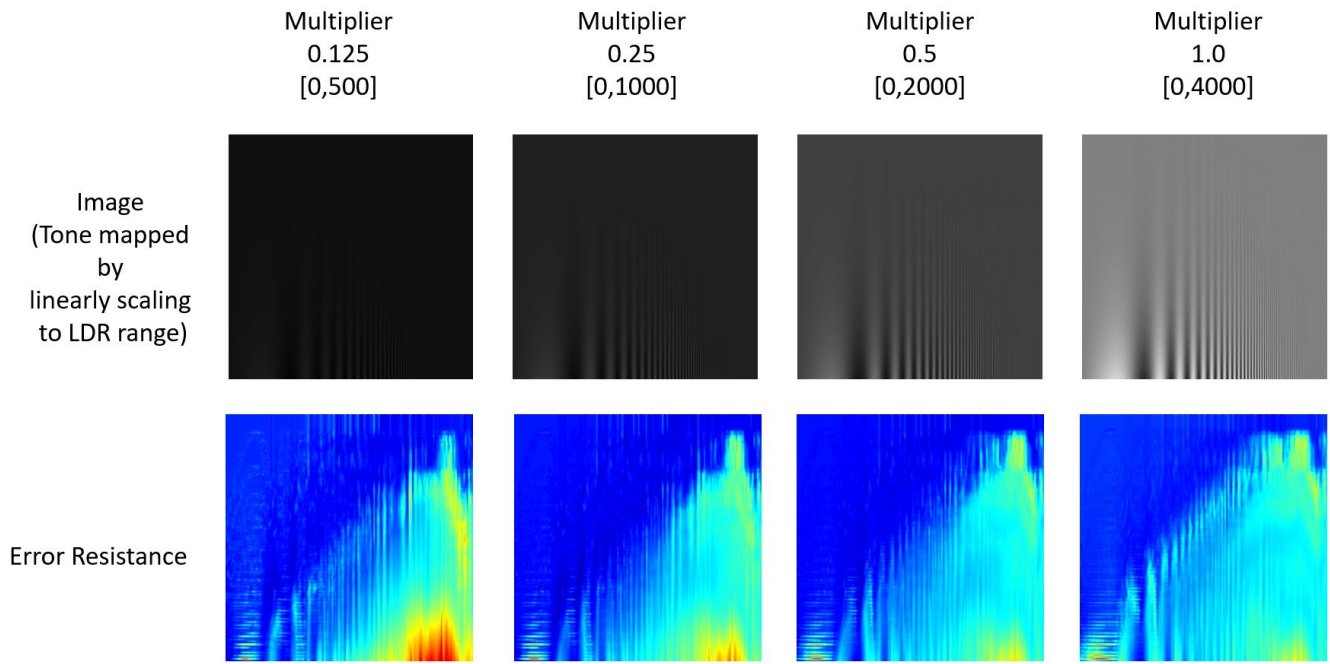

Fig. 9: Input image (top row) and error resistance(bottom row) for different luminance ranges obtained by linearly scaling right top image with starting range [0,4000].

error resistance image.

Another observation here is that the error resistances value is in accordance with the studies done in [1] explained in Section II.B. When the overall intensity is less, the error resistances seem to be higher (more red pixels in second row as we move towards the left), indicating a lower sensitivity to noise. As luminaces get larger, the error resistances decrease and it becomes easier to see noise, indicated by the bluer pixels in second row towards the right.

### D. Error maps

One of the advantages of a CNN based NRIQA scheme is that it gives an error map corresponding to image distortions in addition to the quality score of the image. However, a direct comparison of error maps produced by different schemes would not be informative since different schemes compute final quality differently and will have a different scaling of the error values. PU-SSIM and HDR VDP uses minkovski summation, where as we use a mean value for final quality. However a relative comparison would be helpful to know which areas of the image shows error. The results we report here are on dataset #4 and #5 after the real world test.

A comparison of the error maps produced by HDRVDP, PU-SSIM and Proposed method is shown in fig 1. In the figure, for HDRVDP, the probability of error detection is shown; for PU-SSIM, an inverted SSIM map is shown to show areas with error as high values. Output of proposed algorithm is shown in fourth column. We normalize all the values so that the maximum value is one so as to get a good relative comparison. The images are color coded such that red represents high value, green intermediate values and blue represents a low value.

We choose a range of MOS values to show that the algorithms works in all conditions. It is clear that the values produced agree with the highest performing full reference metric in terms of the location of and the relative intensity of the visual errors.

### E. Failure cases

We found that the algorithm does not produce the correct results in all the cases. One such example is shown in fig 10.

Here we show the error maps for proposed method, HDR-VDP and PU-SSIM and the corresponding error estimation and error resistance images produced using our method. Interpreting the error resistance map is hard since the mixing function has the tunable parameter $k$ that scale down the effect of the error resistance. This can be seen for the case of MOS score 45 in the first row of Fig 10. Minute changes in error resistances produces significant changes in the Error visibility map.

For a MOS value of 74, it is seen that the error maps produced by out method does not correspond to the ones produced by HDRVDP and PUSSIM. This is especially true in the sky region where both full reference algorithm claim that there is minimal distortion, however, our method predicts a higher error. Hence the method seems to have an internal bias that produce error values that tend to be always high in smooth regions.

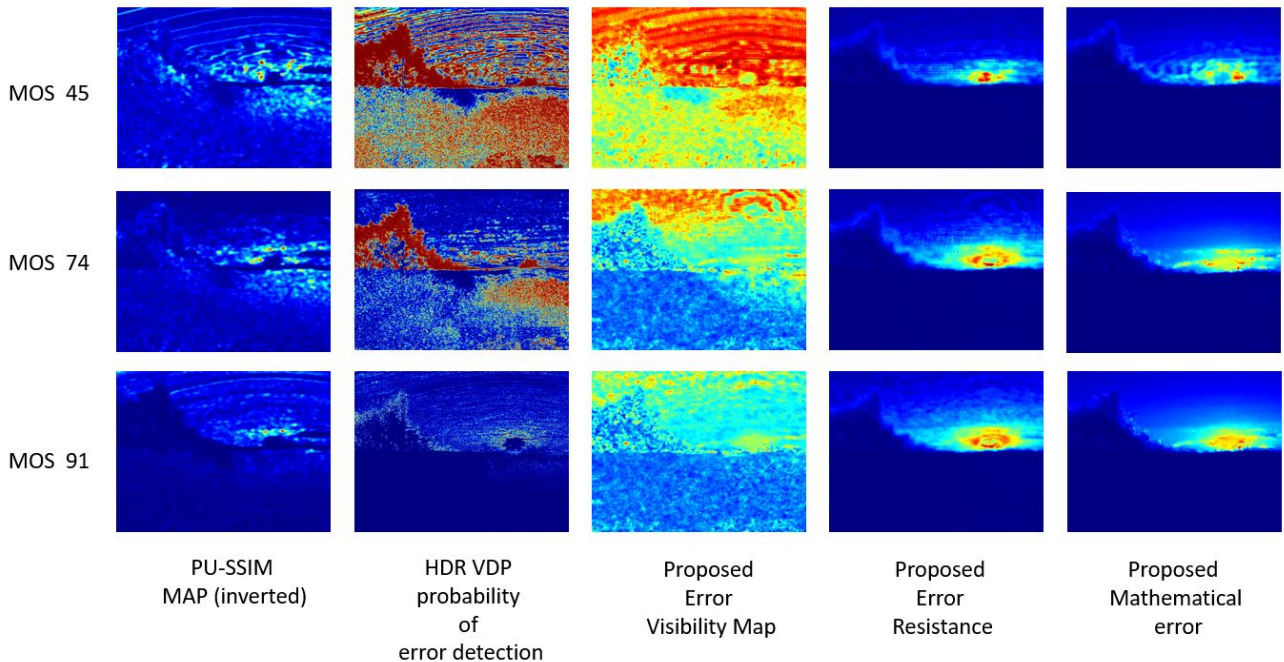

Fig. 10: Comparison of error maps for different MOS values of a chosen image.

## VI. CONCLUSION

We conducted an analysis of LDR-NRIQA for HDR data and discovered that PU encoding is the best way to incorporate quality assessment algorithms developed for LDR out of the box on to HDR data. We then propose a HDR-NRIQA scheme. The scheme uses a convolutional neural network based architecture to generate values corresponding to perceptual effects and true error present in the image and combines it in tunable mixing function. The algorithm predicts the visual distortions in the image because of low level distortion such as compression artifacts. It was found that the the algorithm scores correlate well to human scores and it outperforms state-of-the-art NR IQA methods and is very competitive when compared to HDR FRIQA methods.